\documentclass[journal=jacsat,manuscript=article]{achemso}
\setkeys{acs}{articletitle = true,chaptertitle = true, etalmode=truncate,maxauthors=0}

\usepackage[T1]{fontenc} % Use modern font encodings
\usepackage[english]{babel} %The last language is the default
\usepackage[utf8]{inputenc}

\usepackage{braket}
\usepackage{bm}
\usepackage{siunitx}

\usepackage{booktabs}
\usepackage{threeparttable}
\usepackage{multirow}

\newcommand*{\citen}[1]{%
  \begingroup
    \romannumeral-`\x % remove space at the beginning of \setcitestyle
    \setcitestyle{numbers}%
    \cite{#1}%
  \endgroup   
}

\author{Jil Ludovicy}

\author{Kaveh Haghighi Mood}

\author{Arne Lüchow}
\affiliation[RWTH Aachen University]
{Institute of Physical Chemistry, RWTH Aachen University, Landoltweg 2, 52062 Aachen, Germany}
\email{luechow@rwth-aachen.de}
\phone{+49 241 80 94748}

\title[]
  {Full Wave Function Optimization with Quantum Monte Carlo - A study of the Dissociation Energies of ZnO, FeO, FeH, and CrS}

\begin{document}

\begin{abstract}
The dissociation energies of four transition metal dimers are determined using diffusion Monte Carlo. The Jastrow, CI, and molecular orbital parameters of the wave function are both partially and fully optimized with respect to the variational energy. The pivotal role is thereby ascribable to the optimization of the molecular orbital parameters of a complete active space wave function in the presence of a Jastrow correlation function. Excellent results are obtained for ZnO, FeO, FeH, and CrS. In addition, potential energy curves are computed for the first three compounds at multi-reference diffusion Monte Carlo (MR-DMC) level, from which spectroscopic constants such as the equilibrium bond distance, the harmonic frequency, and the anharmonicity are extracted. All of those quantities agree well with the experiment. Furthermore, it is shown for CrS that a restricted active space calculation can yield improved initial orbitals by including single and double excitations from the original active space into a set of virtual orbitals. We demonstrated in this study that the fixed-node error in DMC can be systematically reduced for multi-reference systems by orbital optimization in compact active spaces. While DMC calculations with a large number of determinants are possible and very accurate, our results demonstrate that compact wave functions may be sufficient in order to obtain accurate nodal surfaces, which determine the accuracy of DMC, even in the case of transition metal compounds.

\end{abstract}

\section{Introduction}
Transition metals and compounds thereof constitute chemical systems that are of great interest in catalytic processes, in electrochemistry as well as in biochemistry. \cite{Stiefel1996,Xiao2016,Kunkes2015,Eshghi2008} Transition metals are highly interesting systems since they compromise open $d$-shells, they exhibit several oxidation states, and they often show magnetic properties. 

The precise understanding of how catalysts work is of major importance when it comes to elucidating and predicting catalytic processes. The bond-breaking between transition metals and main group elements  plays the primary role in these processes,  which justifies the study of transition metal compounds by means of high-level quantum chemical methods. Although the ultimate goal may be to analyze catalytic reactions and other bulk material properties, the investigation of transition metal dimers is the first important step toward an accurate theoretical description of bond breaking processes in bulk materials. The study of these small compounds already poses a considerable challenge due to their strong correlation and complicated electronic structures, thus leading to the calculation of these systems proving rather complex. For the late transition metals, being of great interest especially for catalysis, the static electron correlation yields an important contribution to their properties. The development of efficient methods that are able to capture this part of the correlation thus constitutes a  highly active field of research.  Progress has been made by designing suitable density functional theory (DFT) functionals \cite{Zhao2006,Fuchs2005} as well as by further developing multi-reference wave function based methods, such as multi-reference coupled cluster (MRCC)\cite{Musia2011}.

The use of quantum Monte Carlo (QMC) methods is justified because not only are they highly parallelizable but also because they scale, in general, favorably with the number of electrons.\cite{Luchow2011,Luchow2010} 
It remains a challenge to retain the low-order polynomial scaling in QMC when accounting
for static correlation. 
The variational (VMC) and diffusion (DMC) quantum Monte Carlo approaches  \cite{Austin2012,Wagner2014,Foulkes2001}  are the most widely spread stochastic methods applied in chemistry and physics to determine the properties of systems when an accuracy beyond mean-field theory is required. Transition metal compounds have  successfully been studied with VMC and DMC in the past. Wagner and Mitas \cite{Wagner2007-full} investigated transition metal oxides with fixed-node DMC (FN-DMC). Petz and Lüchow \cite{Petz2011} reported accurate dissociation energies and ionization potentials for sulfide compounds  with FN-DMC. Diedrich \textit{et al.} \cite{Diedrich2005} studied transition metal carbonyls with regard to their dissociation energies with FN-DMC. Horváthová \textit{et al.} \cite{Horvathova2013} presented energetics for transition metal organometallics employing QMC methods. Recently, Doblhoff-Dier \textit{et al.}\cite{Doblhoff-Dier2016} published dissociation energies of 3$d$ transition metal compounds calculated with DMC.

However, for several transition metal compounds, a rather large discrepancy between the theoretical and the experimental dissociation energies can be observed. These compounds are believed to exhibit prominent multi-reference character, the single-determinant approach is not capable of describing these systems correctly. Recently, two of the present authors have shown that a multi-reference ansatz in combination with the optimization of the orbital parameters is necessary to predict the right ground state for and to reproduce the dissociation energy of FeS. \cite{Haghighi2017} In their study, the authors found that the re-optimization of the molecular orbitals (MOs) in the presence of a Jastrow correlation function was the key to obtaining accurate results. 
This multi-reference DMC (MR-DMC) approach with optimization of the orbital parameters is therefore applied to systems for which the single-reference FN-DMC results show significant deviations to the experimental dissociation energies.

Caffarel and coworkers introduced the use of CIPSI wave functions (configuration interaction using perturbative selection made iteratively) as guide functions in DMC. \cite{Giner2013} CIPSI wave functions are large but efficient approximations of the Full-CI wave function, where the dynamic correlation is described through an expansion into determinants. The CIPSI approach was applied to the computation of the F$_2$ potential energy curve\cite{Giner2015} as well as to the calculation of atomic systems\cite{Giner2013,Scemama2014}. In both cases, impressive results were obtained. Scemama \textit{et al.}\cite{Scemama2017} recently applied the CISPI-DMC approach to FeS and confirmed our results.

In contrast to the CIPSI-DMC method, we attempt, in our approach, to describe the systems by a small, physically motivated CAS with both orbitals and CI parameters optimized in the presence of a Jastrow correlation function, which accurately describes the dynamic correlation. 
The accuracy of the DMC method is determined by the accuracy of the nodal surface of the trial wave function. When selected CI wave functions, such as CIPSI, are used, the trial wave function is found to converge toward the Full-CI solution and thus toward the exact nodes (within the basis set limit). The determinant selection is energy-based and it is thus not necessarily optimal for improving the nodal surface. 

With our approach, we explore the possibility to obtain sufficiently accurate DMC energies by retaining compact trial wave functions and by varying the nodal surface through orbital and CI coefficients optimization in the presence of a Jastrow factor. 
While this is also an energy-based optimization and not a direct optimization of the nodal surface, the wave function, and thus the nodal surface, has additional flexibility through the optimization of all parameters. Although the restriction to a compact active space will not allow for accurate total energies, the contributions of higher excitations to the nodal surface may not be important for the
calculation of dissociation energies or other energy differences.

In this paper, ZnO, FeO, FeH as well as CrS are reexamined. Truhlar and coworkers \cite{Xu2015} performed DFT and coupled cluster (CC) calculations for a set of 20 transition metal dimers. They found a significant discrepancy to the experimental dissociation energies of ZnO and FeH for both methods. The authors of that work also state a prominent multi-reference character for those compounds. ZnO and FeH were also examined by Doblhoff-Dier \textit{et al.}\cite{Doblhoff-Dier2016} using DMC with Kohn-Sham (KS) initial orbitals. The significant deviation to the experimental dissociation energies again suggests that a single-reference approach is not able to correctly describe these systems. For FeH, several works\cite{Brown2006,Harrison2008,Shulyak2010} even suggest a breakdown of the Born-Oppenheimer approximation. The investigation of FeO is motivated by our previous work regarding the FeS system. The aim is to obtain as accurate results for FeO as were obtained for the isovalent FeS system. In 2011, one of the current authors presented the evaluation of transition metal sulfides with single-determinant DMC.\cite{Petz2011} The largest deviations were found for FeS and CrS. Therefore, we reexamined CrS with MR-DMC.

\section{Methods}
Since complete reviews on QMC methods as well as on the nature of the wave function ansatz used in this study are available \cite{Hammond1994,Luchow2011,Austin2012}, only a brief overview shall be given in this section. The MR-DMC method employed in this work is thoroughly discussed in our previous publication. \cite{Haghighi2017}

\subsection{Trial Wave Function}
The  trial wave function used in the QMC calculations is of a Slater-Jastrow type,

\begin{equation}
\label{eq:TWF}
\ket{\Psi_\mathrm{T}} = \mathrm{e}^{U} \cdot \sum_i c_i \ket{\Phi_i}
\end{equation}

\noindent with a Jastrow correlation function $\mathrm{e}^{U}$, the configuration interaction (CI) coefficients $c_i$, and the configuration state functions (CSFs) $\ket{\Phi_i}$.  The Jastrow function is totally symmetric with respect to electron permutations and is responsible for the inclusion of the dynamic electron correlation. Throughout this work, a Jastrow factor with electron-electron and electron-nucleus (two-particle terms) as well as with electron-electron-nucleus contributions (three-particle term), developed by Lüchow \textit{et al.}\cite{Luchow2015}, is used. The anti-symmetric part of the trial wave function consists of CSFs which are linear combinations of products of spin-up and spin-down Slater determinants $\ket{D_k^{\uparrow}}$ and $\ket{D_k^{\downarrow}}$, with the coefficients $d_{i,k}$ being determined by the spatial and spin symmetries of the state at hand. 

\begin{equation}
\ket{\Phi_i} = \sum_k d_{i,k} \ket{D_k^{\uparrow}}\ket{D_k^{\downarrow}}
\end{equation}

\noindent The determinants are constructed from molecular orbitals with each MO being expanded into a standard basis set. The trial wave function depends on the coordinates of the electrons as well as on the parameter vector $\mathbf{p} = \{ \bm{\alpha},\mathbf{c},\bm{\kappa} \} $,

\begin{equation}
\ket{\Psi_\mathrm{T}} \equiv \ket{\Psi_\mathrm{T} (\mathbf{R}, \mathbf{p})}
\end{equation}

\noindent with $\bm{\alpha}$ describing the Jastrow parameters, $\mathbf{c}$ being the CI coefficients and $\bm{\kappa}$ corresponding to the orbital rotation parameters. The latter are optimized alternatingly with respect to the VMC energy. There are different methods suitable for energy minimization. Throughout this work, the linear method\cite{Toulouse2007} is used to optimize the Jastrow and CI parameters while the orbital parameters are optimized with the perturbative method (POPT)\cite{Toulouse2007}.  Effective core potentials (ECPs) are used for all calculations in order to include scalar relativistic effects and to decrease the computational demand of QMC by making larger time steps feasible.

\section{Computational Approach}

The trial wave functions were generated with the \textsc{Molpro}\cite{Molpro2} package. The initial wave functions were obtained from \textit{ab initio} calculations, such as HF, KS-DFT (with the B3LYP\cite{Becke1988,Lee1988} functional), and CASSCF. The active space for the latter included the 4\textit{s} and 3\textit{d} orbitals of the metals and the valence \textit{p} orbitals of the main group elements (1\textit{s} for H). The QMC calculations were performed with the program \textsc{Amolqc}\cite{Manten2001,Luchow2014,Luchow2015}, developed in our group. A 69-term Jastrow correlation function (denoted as sm666 in ref. \citen{Luchow2015}) with cusp-less three-particle terms was used for all the calculations. 

All calculations were performed using the ECPs with the triple-$\zeta$ basis sets of Burkatzki, Filippi and Dolg\cite{Burkatzki2007,Burkatzki2008}, referred to in the following by BFD-VTZ. The non-local part of the pseudopotentials was localized on a spherical grid by using the trial wave function. \cite{Fahy1990,Mitas1991}

The dissociation energies were spin-orbit (SO) corrected, and the core-valence (CV) correlation contribution was added. The first-order SO corrections for the atoms were derived from experimental splittings. \cite{NIST} The ones for the molecules were taken from literature. The CV correlation was estimated by means of multi-reference perturbation theory (MR-MP2)\cite{Nakano1993}, as implemented in \textsc{GAMESS}\cite{Schmidt1993}. The core-valence basis set TK+NOSeC-V-QZP with all diffuse functions\cite{Koga1999,Noro2000} was used for these calculations. The dissociation energy was calculated with and without correlating the core electrons to estimate that quantity. The same active space as for the QMC calculations was chosen. The importance of including the core-valence correlation contribution in order to obtain accurate transition metal properties was also noted in other studies. \cite{Aoto2017}
The zero-point energy (ZPE) of CrS was determined with the electron structure modelling program \textsc{Gaussian}\cite{Gaussian} at B3LYP/BFD-VTZ\cite{Burkatzki2007,Burkatzki2008} level. The ZPEs of the other compounds were obtained from a Morse fit to the MR-DMC potential energy curves, computed with fully optimized guide functions.

\section{Results and Discussion}

In this section, the single- and multi-determinant QMC calculations for different transition metal dimers will be discussed. The effect of the MO optimization on the VMC and DMC energies for both approaches is to be investigated in particular. Finally, the dissociation energies of the different compounds are evaluated. The absolute energies will only be discussed for the first system, the ones of the remaining compounds can be found in the supporting information. The DMC energies of the atomic species are given there as well.

Elementary information about the dimers is summarized in Table \ref{tab:general_Info}. Further details are provided in the sections describing the different compounds.
The CSF column in Table \ref{tab:general_Info} represents the number of CSFs with non-zero coefficients, necessary to describe the ground state of the respective compound for a given CAS.

\begin{table}[H]
\centering
\caption{Ground state, active space, CSFs, equilibrium bond distance (\si{\angstrom}), spin-orbit correction, core-valence correlation contribution, and zero point energy for all four compounds. The energy corrections are given in eV.}
\label{tab:general_Info}
\begin{threeparttable}
\begin{tabular}{cccccccc}
\toprule
Compound & Ground state & CAS & CSFs & $r_{e}$  & SO   & CV  & ZPE   \\
\midrule
ZnO & $^{1}\Sigma^{+}$\cite{Fancher1998,Bauschilcher1998} & [16,9] & 10 & 1.709 & n/a &  0.0923 & 0.0462 \\
FeO & $^{5}\Delta$\cite{Cheung1982,Drechsler1997} & [12,9] & 184 & 1.623 & -0.0558 \cite{Schultz2005} & 0.126 & 0.0537 \\
FeH & $^4\Delta$\cite{Stevens1983,Schultz1991} & [9,7] & 30 & 1.567 & -0.0477\cite{Xu2015} & 0.0675 & 0.114 \\
CrS & $^5\Pi$\cite{Shi2001,Pulliam2010} & [10,9] & 670 & 2.0781\cite{Pulliam2010} & -0.0118\cite{Pulliam2010} & 0.140 & 0.0278\tnote{a} \\
\bottomrule
\end{tabular} 
\begin{tablenotes}
\item[a] calculated at B3LYP/BFD-VTZ level.
\end{tablenotes}
\end{threeparttable}
\end{table}

	\subsection{ZnO}
	
ZnO exhibits an electronic configuration of $\pi^{4}{\sigma^2}\delta^{4}\pi^{4}\sigma^{2}$. The [16,9]-CAS is built from the 4\textit{s} and 3\textit{d} orbitals of zinc, and from the 2\textit{p} orbitals of oxygen. The equilibrium bond distance, see Table \ref{tab:general_Info}, is obtained from a potential curve at MR-DMC level. The curve was recorded for a small time step of $\tau = 0.0005$ a.u. in order to reach an acceptance ratio of about 99\%. The different wave function parameters were optimized alternatingly with respect to the VMC energies. The latter are illustrated in Table \ref{tab:ZnO-energies} together with the zero time step extrapolated DMC energies. The non-optimized parameters, such as the CI and the MO coefficients, are taken from the respective \textit{ab initio} calculations for some optimization levels.

\begin{table}[H]
\centering
\caption{ZnO VMC and DMC energies in $E_\mathrm{h}$ at various optimization levels (Jas = Jastrow only), using different starting orbitals and BFD-VTZ/sm666.}
\label{tab:ZnO-energies}
\begin{tabular}{ccccc}
\toprule
Ansatz &  Orbitals &Optimization level & VMC energy  & DMC energy \\ 
\midrule \midrule
\multirow{3}*{Single det}& HF & Jas & -242.8836(3) & -242.9931(5)  \\ 
 
& B3LYP & Jas & -242.8944(3) & -243.0022(5)  \\ 
 
& opt   & Jas+MO &  -242.9013(3) & -243.0065(6) \\  
 \midrule 
  \multirow{3}*{[16,9]-CAS} & CAS & Jas & -242.8971(3) & -242.9950(5)  \\
  
  & CAS & Jas+CI & -242.9047(3) & -243.0023(6) \\ 
 
   & opt  & Jas+MO+CI & -242.9176(3) & -243.0111(5) \\ 
\bottomrule
\end{tabular} 
\end{table}	

A systematic lowering of the VMC energies can be observed from HF over KS (B3LYP) to CAS orbitals. The MO optimization leads to an improvement of the energies in both cases, with the change being more significant for the CAS orbitals. It is interesting to see that the coupling between dynamic and static correlation, which will only be included by optimizing the orbitals in the presence of the Jastrow correlation function, has a substantial impact on the energies. The use of KS orbitals in QMC, on the other hand, partly captures this effect. The fully optimized wave function thus captures the dynamic correlation explicitly through the Jastrow factor while the static one is included through the different configurations.

The optimization of the molecular orbital parameters in the presence of a Jastrow correlation function leads to a substantial improvement of the nodal surface for the CAS ansatz. For the KS orbitals, the DMC energy is only slightly lowered by the optimization of the MO parameters which emphasizes that the nodal surface was already almost optimal for a single-determinant ansatz before the optimization. The optimization of all parameters (Jas, MO, and CI) is necessary to obtain DMC energies that are lower than the ones with the optimized KS orbitals. This reveals the influence of the dynamic correlation on the quality of the nodal surface. Without optimizing the MO and CI parameters of the CAS initial wave function, the Jastrow optimization does not change the nodal surface. The VMC as well as the DMC energies follow similar trends for all dimers and are thus only discussed for ZnO.

Table \ref{tab:DE-ZnO} contains the dissociation energies of ZnO at different optimization levels. As for the atomic species, the DMC energies for the different starting orbitals (and optimization levels) only differed within the order of the statistical error. The core-valence correlation contribution of ZnO amounts to 0.0923 eV. It yields a substantial contribution to the dissociation energy, making it  a non-negligible quantity. ZnO does not exhibit first-order spin-orbit coupling due to its totally symmetric ground state. The ZPE was determined by means of the potential energy curve and corresponds to 0.0462 eV.

\begin{table}[H]
\centering
\caption{DMC dissociation energies of ZnO in eV at various optimization levels, using different starting orbitals and BFD-VTZ/sm666. The energies are CV and SO corrected.}
\label{tab:DE-ZnO}
\begin{tabular}{cccc}
\toprule
 Ansatz & Orbitals  & Optimization level & $D_0$  \\ 
\midrule \midrule
\multirow{3}*{Single det} & HF & Jas & 1.20(2)  \\ 
 
& B3LYP & Jas & 1.45(2)  \\ 
 
& opt    & Jas+MO & 1.57(2)  \\  
    \midrule
 
  \multirow{3}*{CAS} & CAS & Jas & 1.25(2)  \\ 
  
 & CAS 	& Jas+CI & 1.45(2)  \\
 
 & opt    & Jas+MO+CI & 1.69(2) \\ 
\bottomrule
\end{tabular} 
\end{table}

\begin{table}[H]
\centering
\caption{Bond dissociation energies in eV calculated and measured for ZnO.}
\label{tab:DE-ZnO-literature}
\begin{tabular}{lccc}
\toprule
Investigators & Method & $D_\mathrm{e}$ & $D_0$ \\ 
\midrule \midrule
This work & SR-DMC & 1.61(2) & 1.57(2) \\
This work & MR-DMC & 1.74(2) & 1.69(2) \\
\midrule
Clemmer \textit{et al.}\cite{Clemmer1991} & Mass Spectrometry &  & 1.61(4) \\ 
Zhang \textit{et al.}\cite{Zhang2013a} & from $\Delta H_\mathrm{f}$ & 1.65(4) &  \\ 
 \midrule
 Krogel \textit{et al.}\cite{Krogel2016} & DMC & 1.45(2) & \\
Weaver \textit{et al.}\cite{Weaver2013} & CASPT2 & 1.54 & 1.45 \\ 
 Aoto \textit{et al.}\cite{Aoto2017} & CCSD(T)(CV)/CBS-DK & 1.55  &  \\
 \multirow{3}*{Xu \textit{et al.}\cite{Xu2015}} & CCSDT(2)$_\mathrm{Q}$/apTZ-DK(3) & 1.36 & \\
  & DFT/B97-1-DK & 1.30 & \\
  & DFT/M06-L-DK & 1.35 & \\
\bottomrule 
\end{tabular} 
\end{table}

\begin{figure}[H]
\begin{center}
  \includegraphics[scale=0.8]{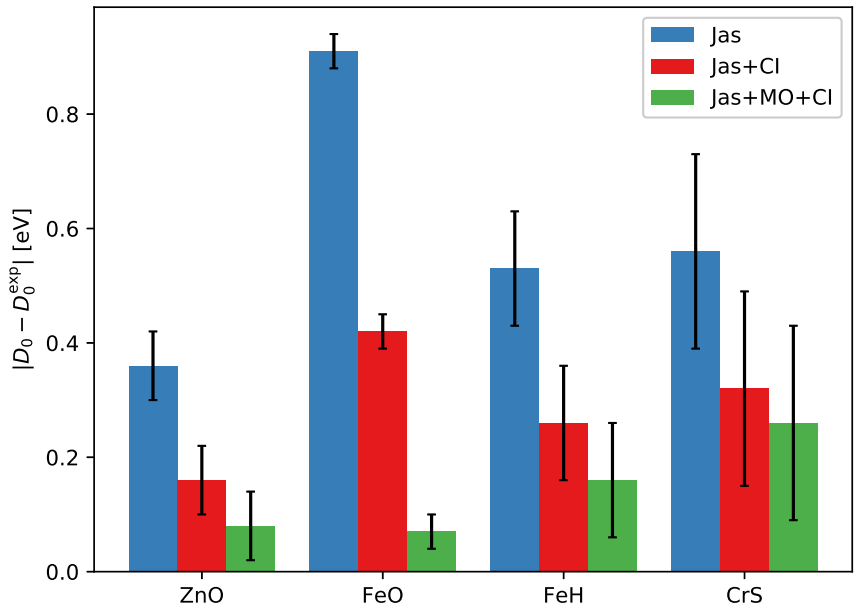}
\end{center}
\vspace{-1cm}
\caption{Deviation to experimental dissociation energies at the various optimization levels for all four compounds.}
\label{fig:Deviation}
\end{figure}

The DMC ansatz with HF nodes fails to reproduce the dissociation energy of ZnO, see Table \ref{tab:DE-ZnO-literature}. At the Jas optimization level, the KS orbitals yield a significantly more accurate dissociation energy than the HF and CAS nodes. Without the MO optimization, the DFT and the CAS guide functions yield comparable results. The optimization of the orbital parameters in the presence of a Jastrow correlation function improves the dissociation energy substantially.  
In comparison to the experimental dissociation energy\cite{Clemmer1991,Zhang2013a}, 
the single determinant dissociation energy is improved by more than 0.1 eV when optimizing the B3LYP orbitals together with the Jastrow function. The resulting dissociation energy is accurate to less 
than 0.1 eV. 
The fully optimized MR-DMC dissociation energy is slightly less accurate but still better than 0.1 eV.
The significant improvement of the dissociation energy for the MR-DMC approach by optimizing not only 
the Jastrow parameters but also the CI and the MO parameters is visualized in Fig. \ref{fig:Deviation}.

A single-reference treatment thus seems to be accurate enough for the ZnO system. This is in contrast to the findings by Xu \textit{et al.}\cite{Xu2015} based on different multi-reference diagnostics and the discrepancy they found in comparison with the experimental dissociation energies for CC as well as for DFT. Doblhoff-Dier \textit{et al.}\cite{Doblhoff-Dier2016} also reported a large deviation to the experiment for their single-reference DMC dissociation energies (about 0.3 eV for ZnO), using KS determinants with different functionals. We believe however, that the discrepancy of our SR-DMC dissociation energy to this work is due to the orbital optimization and the core-valence correlation contribution to the dissociation energy.
 
Krogel \textit{et al.}\cite{Krogel2016} published single-determinant DMC potential energy curves of transition metal oxides. Their dissociation energy for ZnO is significantly lower than the one obtained in this work with SR-DMC. The CASPT2 method is not able to reproduce the experimental dissociation energy of Clemmer and coworkers.  The CC dissociation energy of Aoto \textit{et al.}\cite{Aoto2017}, extrapolated to the complete basis set, core-valence corrected and including scalar relativistic effects underestimates the experimental dissociation energy by about 0.1 eV. Xu \textit{et al.}\cite{Xu2015} studied the performance of CC compared to DFT calculations. Their CC and DFT dissociation energies are significantly lower than the one computed in this work. 

	\subsection{FeO}
	
The ground state of FeO corresponds to an electronic configuration of 	$\sigma^2\pi^4\sigma^2\delta^3\sigma^1\pi^2$. The potential energy curve at MR-DMC level is computed with a fixed time step of $\tau= 0.001$ a.u. The spin-orbit correction of -0.0558 eV, see Table \ref{tab:general_Info}, is taken from ref. \citen{Schultz2005}, the authors of which made use of an earlier experimental study.\cite{Merer1989} The wave function is fully optimized with an sm666 Jastrow factor. The VMC and DMC energies of FeO are listed in the supporting information. They are not discussed since they show trends comparable to the ones that were already observed for the ZnO compound.

\begin{table}[H]
\centering
\caption{DMC dissociation energies of FeO in eV at various optimization levels, using different starting orbitals and BFD-VTZ/sm666. The energies are CV and SO corrected.  }
\label{tab:DE-FeO}
\begin{tabular}{cccc}
\toprule
 Ansatz & Orbitals  & Optimization level &  $D_0$ \\ 
\midrule \midrule
\multirow{3}*{Single det} & HF & Jas & 2.88(2)   \\ 
 
& B3LYP & Jas & 3.69(2)  \\ 
 
& opt   & Jas+MO & 3.83(2) \\  
    \midrule
 
  \multirow{3}*{CAS} & CAS & Jas & 3.27(2) \\ 
 
&CAS 	& Jas+CI & 3.76(2) \\ 
& opt     & Jas+MO+CI & 4.11(2) \\ 
\bottomrule
\end{tabular} 
\end{table}

The dissociation energies of FeO are listed in Table \ref{tab:DE-FeO}. Similarly to the ZnO compound, the optimization of the MO parameters improves the dissociation energy to a large extent. The approach used in this work is able to systematically improve the dissociation energy for the different wave function ansätze and optimization levels, see Figure \ref{fig:Deviation}. The single-determinant as well as the  CAS guide functions without MO optimization yield similar results, they all underestimate the dissociation energy substantially. Only the dissociation energy of the fully optimized multi-reference guide function is in good agreement with the experimental results, realized by means of various methods\cite{Chestakov2005, Li2009, Smoes1984}  (cf. Table \ref{tab:DE-FeO-literature}).

\begin{table}[H]
\centering
\caption{Bond dissociation energies in eV calculated and measured for FeO.}
\label{tab:DE-FeO-literature}
\begin{tabular}{lccc}
\toprule
Investigators & Method & $D_\mathrm{e}$ &  $D_0$  \\ 
\midrule \midrule
This work & MR-DMC   & 4.17(2) &  4.11(2) \\
\midrule
Chestakov \textit{et al.}\cite{Chestakov2005} & Photodissociation & &   4.18(1) \\ 

Li \textit{et al.}\cite{Li2009} & Collision-induced dissociation & & 4.18(1)
  \\ 
  
 Smoes and Drowart \cite{Smoes1984} & Mass spectrometry & & 4.16(8) \\ 
 \midrule
Krogel \textit{et al.}\cite{Krogel2016} & DMC & 4.25(1) & \\ 
 
Aoto \textit{et al.}\cite{Aoto2017} & CCSD(T) & 4.21 & 
 \\ 
 
 Sakellaris \textit{et al.} \cite{Sakellaris2011} & MRCI+DKH2+Q & 3.69 &  \\

\multirow{3}*{Jensen \textit{et al.}\cite{Jensen2007-DFT}} &  DFT/B3LYP & & 3.96 \\

& DFT/BP86 & & 5.21  \\

& DFT/PBE & & 5.31 \\
 
\bottomrule 
\end{tabular} 
\end{table}

The  DMC approach of Krogel \textit{et al.}\cite{Krogel2016} yields a dissociation energy that is comparable to the one computed in this study. The CC dissociation energy of Aoto \textit{et al.}\cite{Aoto2017} agrees well with the experimental ones as well as with the MR-DMC value of this work, while the MRCI result of Sakellaris \textit{et al.}\cite{Sakellaris2011} is significantly lower. As for the DFT approaches of Jensen \textit{et al.}\cite{Jensen2007-DFT}, the different functionals are not able to yield consistent results, either severely under- or overestimating the experimental dissociation energies.

	\subsection{FeH}
	
The ground state of FeH is described by the electron configuration $\sigma^2\pi^4\delta^2\sigma^1$. The [9,7]-CAS, see Table \ref{tab:general_Info}, is constructed from the 4\textit{s} and 3\textit{d} orbitals of iron, and the 1\textit{s} orbital of hydrogen. The MR-DMC potential energy curve is recorded at a fixed time step of $\tau = 0.001$ a.u.

Table \ref{tab:DE-FeH} shows a significantly higher dissociation energy for the KS nodes than for the HF nodes at the Jastrow optimization level. The optimization of the molecular orbital parameters shows no improvement of the dissociation energy for the single determinant guide function. For the multi-reference approach, a systematic improvement of the dissociation energy can be observed for the different optimization levels, which is visualized by Figure \ref{fig:Deviation}. The MO optimization of the CAS guide function in the presence of a Jastrow correlation factor has a significant effect on the dissociation energy of FeH, it is increased by about 0.4 eV. Note that even if the anti-symmetric part of the multi-reference guide function is not optimized, a more accurate dissociation energy is obtained compared to the one from the fully optimized single-reference wave function, which speaks for FeH exhibiting a prominent multi-reference character.

\begin{table}[H]
\centering
\caption{DMC dissociation energies  of FeH in eV at various optimization levels, using different starting orbitals and BFD-VTZ/sm666. The energies are CV and SO corrected.}
\label{tab:DE-FeH}
\begin{tabular}{cccc}
\toprule
 Ansatz & Orbitals  & Optimization level & $D_0$  \\ 
\midrule \midrule
\multirow{3}*{Single det} &HF & Jas & 0.81(2)  \\ 

& B3LYP & Jas & 1.02(2)  \\ 
 
& opt  & Jas+MO & 1.02(2)  \\  
    \midrule
 
  \multirow{3}*{CAS} & CAS & Jas & 1.10(2)  \\ 
 & CAS	& Jas+CI & 1.37(2)   \\
 & opt    & Jas+MO+CI & 1.79(2)  \\ 
\bottomrule
\end{tabular} 
\end{table}	

A good agreement is achieved when comparing the MR-DMC dissociation energy to the experimental one of Schultz and Armentrout \cite{Schultz1991}, see Table \ref{tab:DE-FeH-literature}. The breakdown of the BO approximation, which was mentioned in several studies\cite{Brown2006,Harrison2008,Shulyak2010}, can be refuted by the accurate MR-DMC results. The DFT calculations with different functionals, performed by Jensen \textit{et al.}\cite{Jensen2007-DFT}, fail to yield satisfactory results since they severely overestimate the dissociation energy. The CC results of Aoto \textit{et al.}\cite{Aoto2017} and Cheng \textit{et al.}\cite{Cheng2017} agree well with our dissociation energy. The focal point analysis (FPA) of DeYonker and Allen\cite{DeYonker2012} yields a substantial deviation to the experimental dissociation energy.
Xu and coworkers \cite{Xu2015} confirmed the multi-reference character of FeH by different diagnostics. 
However, they obtained an accurate dissociation energy with CCSDT(2)$_\mathrm{Q}$ including scalar relativistic effects, while their reported DFT dissociation energies are severely larger than the experimental value.
Nonetheless, they argued that KS DFT yields overall comparable results to CC theory for the twenty transition metal compounds that they investigated.

\begin{table}[H]
\centering
\caption{Bond dissociation energies in eV calculated and measured for FeH.}
\label{tab:DE-FeH-literature}
\begin{tabular}{lccc}
\toprule
Investigators & Method & $D_\mathrm{e}$ & $D_0$ \\ 
\midrule \midrule
This work & MR-DMC & 1.90(2) & 1.79(2) \\
\midrule
Schultz and Armentrout\cite{Schultz1991} & Mass Spectrometry &  & 1.63(8) \\

 \midrule
 
\multirow{3}*{Jensen \textit{et al.}\cite{Jensen2007-DFT}} & DFT/B3LYP & 2.10

& \\

& DFT/BP86 & 2.41  & \\ 

& DFT/PBE & 2.30 & \\
 \midrule
 \multirow{3}*{Xu \textit{et al.}\cite{Xu2015}} & CCSDT(2)$_\mathrm{Q}$/apTZ-DK(3) & 1.78 & \\
   & DFT/B97-1-DK &  2.00 & \\
  & DFT/M06-L-DK & 2.17  & \\
Aoto \textit{et al.}\cite{Aoto2017} & CCSD(T)(CV)/CBS-DK & 1.95  &  \\ 
Cheng \textit{et al.}\cite{Cheng2017} & CCSD(T) & 1.99 & \\ 

DeYonker and Allen \cite{DeYonker2012} & FPA &  & 1.86 \\ 
\bottomrule 
\end{tabular} 
\end{table}

	\subsection{CrS}
	\label{subsec:CrS}

A slightly modified initial wave function compared to the usual CAS ansatz is chosen for the CrS system due to the inability to converge the MO parameters with QMC when starting from CAS orbitals generated by a [10,9]-CASSCF calculation. The active orbitals of the CAS wave function are further relaxed by performing a RASSCF calculation with single and double excitations into a set of virtual orbitals. The original [10,9]-CAS corresponds to the RAS2, where all possible excitations are performed while a RAS3 with 11 virtual orbitals is created for the single and double excitations from the RAS2. The RAS1 remains empty. The RASSCF calculation will henceforth be referred to as [10,9;2,11]-RAS calculation. The aim of this approach is to obtain better initial orbitals that can then be further optimized in a QMC energy minimization calculation. The CAS (=RAS2) orbitals are optimized in the partial presence of dynamic correlation through excitations to the RAS3. The RAS2 orbitals are hence expected to be closer to the converged orbitals in a full MR-VMC optimization. The CAS for the QMC calculations is however built similarly to the one of the other compounds, namely from the 4\textit{s} and the 3\textit{d} orbitals of chromium, and from the 3\textit{p} orbitals of sulfide. 

The ground state of CrS is described by the electron configuration $\sigma^2\pi^4\sigma^1\delta^2\pi^1$. The calculations for CrS are performed with the experimental bond length of 2.0781 \si{\angstrom}.\cite{Pulliam2010}

\begin{table}[H]
\centering
\caption{DMC dissociation energies of CrS in eV at various optimization levels, using different starting orbitals and BFD-VTZ/sm666. The energies are CV and SO corrected.}
\label{tab:DE-CrS}
\begin{tabular}{cccc}
\toprule
  Ansatz & Orbitals  & Optimization level & $D_0$ \\ 
\midrule \midrule
\multirow{3}*{Single det} &HF & Jas & 2.05(2) \\ 

& B3LYP & Jas & 2.77(2) \\ 
 
&  opt & Jas+MO & 2.77(2) \\  
    \midrule
    \multirow{5}*{CAS} & CAS & Jas & 2.43(2) \\
   
    & CAS & Jas+CI &  2.70(2) \\

  & RAS2 & Jas &  2.80(2) \\ 
  
  &RAS2	& Jas+CI & 3.04(2)  \\
 
  & opt   & Jas+MO+CI & 3.10(2) \\  
\bottomrule
\end{tabular} 
\end{table}

The dissociation energies of CrS for the different approaches are listed in Table \ref{tab:DE-CrS}. Similarly to the other compounds, a systematic improvement of the dissociation energy can be observed for the different methods and optimization levels. The KS nodes appear ideal since the MO optimization does not improve the dissociation energy. The ansatz with CAS orbitals yields lower dissociation energies than the one obtained with KS orbitals at the same optimization level. Relaxing the initial active orbitals through a RAS calculation has a substantial effect on the dissociation energy. Not only are the dissociation energies significantly improved when comparing them to the ones obtained with CAS initial orbitals but also the molecular orbital parameters could be successfully optimized. At a given optimization level, the dissociation energies for the different CAS nodes differ by about 0.3 eV. When further optimizing the orbitals initially taken from the RAS calculation, the dissociation energy can be improved by 0.05 eV.  Figure \ref{fig:Deviation} shows that the deviations between the experimental dissociation energies and the ones computed with MR-DMC can be systematically reduced by increasing the level of optimization.

\begin{table}[H]
\centering
\caption{Bond dissociation energies in eV calculated or measured for CrS.}
\label{tab:DE-CrS-literature}
\begin{tabular}{lcc}
\toprule
Investigators & Method & $D_0$  \\ 
\midrule \midrule
This work & MR-DMC &  3.10(2) \\
\midrule
Drowart \textit{et al.}\cite{Drowart1967} &  Mass Spectrometry & 3.36(15) \\
\midrule

Petz and Lüchow\cite{Petz2011} & DMC/PPII & 2.969(9) \\
 
 Bauschlicher and Maitre\cite{Bauschilcher1995} & CCSD(T) & 2.89 \\ 
 
 Liang and Andrews\cite{Liang2002} & DFT/BPW91 & 3.33 \\
 
\bottomrule 
\end{tabular} 
\end{table}

Table \ref{tab:DE-CrS-literature} yields experimental and theoretical dissociation energies for CrS. The calculated dissociation energy is larger than both, the single-determinant DMC\cite{Petz2011} and the CCSD(T)\cite{Bauschilcher1995} values, but still smaller than the experimental $D_0$ of Drowart \textit{et al.}\cite{Drowart1967}. Assessing the accuracy of the MR-DMC result proves challenging due to the large experimental error bar. Our dissociation energy is about 0.1 eV below the lower bound of Drowart and coworkers. In order to estimate the accuracy of the obtained MR-DMC result, experimental data with smaller error bars are needed.

\subsection{Spectroscopic Constants}

The potential energy curves of ZnO, FeO, and FeH were computed at the fully optimized MR-DMC level and fitted to Morse functions from which spectroscopic constants, such as the equilibrium bond distance (minimum of the Morse curve), the harmonic frequency as well as the anharmonicity could be deduced. The evaluation of those constants allows an assessment of the employed method. Table \ref{tab:potE-curve} illustrates the obtained quantities and compares them for different methods. 

For ZnO, the equilibrium bond distance  is in good agreement with the experimental bond length of Zack \textit{et al}. \cite{Zack2009} and it is slightly shorter than the CC and DFT ones.  The MR-DMC bond distance of FeO is slightly larger than the one obtained from other theoretical methods and it agrees with the experiment. As for FeH, the equilibrium bond distance obtained from the Morse fit is similar to the one from other theoretical studies.

The harmonic frequencies and the anharmonicities obtained from the Morse fit are in good agreement as well with the experimental as with the theoretical results for all three compounds.

\begin{table}[H]
\centering
\caption{Spectroscopic constants for the different transition metal compounds. The equilibrium bond distance is given in \si{\angstrom}, the harmonic frequency and the anharmonicity in \si{\per\centi\meter}.}
\label{tab:potE-curve}
\hspace*{-1.6cm}
\begin{threeparttable}
\begin{tabular}{clcccc}
\toprule
System & Investigators & Method & $r_e$  & $\omega_\mathrm{e}$ & $\omega_\mathrm{e} x_\mathrm{e}$  \\ 
\midrule \midrule
\multirow{6}*{ZnO}&This work & MR-DMC & 1.709  &  746(8) & 4.4(1)   \\ 
\cmidrule{2-6}

& Zack \textit{et al.}\cite{Zack2009} & Direct-absorption methods & 1.7047(2) & 738 & 4.88  \\

& Fancher \textit{et al.}\cite{Fancher1998}  & Photoelectron Spectrum  & & 805(40) &  \\ 
\cmidrule{2-6}
& Weaver \textit{et al.}\cite{Weaver2013} & CASPT2 & 1.7 & 742 &  \\ 

& \multirow{2}*{Bauschlicher and Partridge\cite{Bauschilcher1998}} & CCSD(T) & 1.719 & 727.2 & 5.83  \\

& & DFT/B3LYP & 1.713 & 741 &  \\
\midrule\midrule 
\multirow{6}*{FeO} &This work & MR-DMC & 1.623 &  866(79) & 4.7(7)   \\ 
\cmidrule{2-6}
& Allen \textit{et al.}\tnote{a} &  & 1.619 & & \\

&Drechsler \textit{et al.}\cite{Drechsler1997} & anion-ZEKE &  & 882 & 4 \\
\cmidrule{2-6}
& Hendrickx and Anam \cite{Hendrickx2009} & CASPT2 & 1.612 & 887 & \\ 

&\multirow{2}*{Sakellaris \textit{et al.}\cite{Sakellaris2011}}& MRCI & 1.612 & 864 & 7.2 \\
& & RCCSD(T) & 1.607 & 905 & 5.9 \\
\midrule\midrule 
\multirow{5}*{FeH}& This work & MR-DMC & 1.567 &  1842(27) & 38.9(9)   \\ 
\cmidrule{2-6}
&Philips \textit{et al.} \cite{Phillips1987} & Near IR Spectrum  &  & 1826.86 & 31.96 \\
&Dulick \textit{et al.} \cite{Dulick2003} & &  & 1831.8(19) & 34.9(9) \\
\cmidrule{2-6}
&DeYonker and Allen \cite{DeYonker2012} & CCSDT & 1.5660  & 1798.8 & 37.8 \\

&Jensen \textit{et al.} \cite{Jensen2007-DFT} &DFT/B3LYP & 1.57 && \\
\bottomrule 
\end{tabular} 
\begin{tablenotes}
\item[a] derived from Allen \textit{et al.}\cite{Allen1996-FeO}
\end{tablenotes}
\end{threeparttable}
\end{table}

\section{Conclusion}

The dissociation energies of ZnO, FeO, FeH, and CrS were determined through single- and multi-determinant DMC calculations. The Jastrow, CI, and MO parameters of the wave functions were both partially and fully optimized with respect to the energy. A systematic improvement of the dissociation energy could be observed for all compounds for the different ansätze.  In the single determinant approach, optimizing the KS orbitals led for all four systems to either minor or no significant improvement of the nodal surface of the guide functions. For the multi-reference ansatz, on the other hand, the optimization of the molecular orbital parameters in the presence of a Jastrow correlation function is the key contribution. A good agreement of the MR-DMC dissociation energy with the experimental ones was achieved for ZnO, FeO, and FeH. We found that the ZnO dissociation energy could be obtained within 0.1 eV already with a single-reference ansatz, but only after MO optimization. In addition, potential energy curves at MR-DMC level were recorded for these three compounds, which yielded equilibrium bond distances and spectroscopic constants that agree well with literature. As for CrS, the complex MO optimization was tackled by employing more accurate initial orbitals, generated by a RASSCF calculation. The calculated dissociation energy of CrS agrees well with other theoretical methods. Unfortunately, the error bar of the experimental dissociation energy is rather large which is why the accuracy of the obtained result is difficult to assess. Our results show that it is possible to obtain accurate dissociation energies and properties by compact wave functions generated from a small, physically motivated CAS.

\begin{acknowledgement}

The authors thank the Jülich-Aachen research alliance (JARA) and the RWTH Aachen University  for the granted computing time under project rwth0278. The authors thank Christina Zitlau for the performance of preliminary calculations.

\end{acknowledgement}

\begin{suppinfo}

\begin{table}[H]
\centering
\caption{ Ground states, DMC energies (BFD-VTZ/sm666) in $E_\mathrm{h}$, and  spin-orbit (SO) corrections (in eV) of the different atomic species.}
\label{tab:Atoms}
\begin{tabular}{ccccc}
\toprule
Element  & Ground State &  Optimization level & Energy & SO correction \\ 
\midrule 
Zn & $^1$S & Jas+MO & -227.0565(5) & n/a \\

Fe  & $^5$D &  Jas+MO & -123.8126(4) & -0.050\\ 

Cr & $^7$S & Jas+MO & \hphantom{1}-86.9010(4) & n/a \\ 

O  & $^3$P & Jas+MO &  \hphantom{1}-15.8938(1) & -0.010  \\ 

H & $^2$S & / & \hspace{-0.1cm}-0.5000 & n/a \\
 
S   & $^3$P & Jas+MO &  \hphantom{1}-10.1314(1) & -0.024 \\ 

\bottomrule
\end{tabular} 
\end{table}

\begin{table}[H]
\centering
\caption{FeO VMC and DMC energies in $E_\mathrm{h}$ at various optimization levels, using different starting orbitals and BFD-VTZ/sm666.}
\label{tab:FeO-energies}
\begin{tabular}{ccccc}
\toprule
  Ansatz & Orbitals & Optimization level & VMC energy  & DMC energy  \\ 
\midrule \midrule
 \multirow{3}*{Single det} & HF & Jas &  -139.7003(5) & -139.8099(6) \\ 
 
& B3LYP & Jas & -139.7326(4) & -139.8394(6) \\ 
 
 & opt & Jas+MO & -139.7499(4) & -139.8445(6) \\  
 \midrule
 
  \multirow{3}*{[12,9]-CAS} & CAS& Jas & -139.7369(4) & -139.8239(6)  \\ 
& CAS  & Jas+CI & -139.7552(4) & -139.8421(6) \\
 & opt    & Jas+MO+CI & -139.7708(3) & -139.8550(6) \\ 
\bottomrule
\end{tabular} 
\end{table}

\begin{table}[H]
\centering
\caption{FeH VMC and DMC energies in $E_\mathrm{h}$ at various optimization levels, using different starting orbitals and BFD-VTZ/sm666.}
\label{tab:FeH-energies}
\begin{tabular}{ccccc}
\toprule
 Ansatz & Orbitals & Optimization level & VMC energy & DMC energy  \\ 
\midrule \midrule
\multirow{3}*{Single det} & HF & Jas & -124.2815(2) & -124.3443(5)  \\ 

& B3LYP & Jas & -124.2923(2) & -124.3519(5)  \\ 
 
  & opt  & Jas+MO & -124.2948(2)  & -124.3519(5) \\  
 \midrule
  \multirow{3}*{[9,7]-CAS} & CAS & Jas & -124.2940(2) & -124.3548(5)  \\ 
 
 & CAS	& Jas+CI & -124.3030(2) & -124.3647(5) \\
 & opt    & Jas+MO+CI & -124.3252(2) & -124.3802(5) \\ 
\bottomrule
\end{tabular} 
\end{table}

\begin{table}[H]
\centering
\caption{CrS VMC  and DMC energies in $E_\mathrm{h}$ at various optimization levels, using different starting orbitals and BFD-VTZ/sm666.}
\label{tab:CrS-energies}
\begin{tabular}{ccccc}
\toprule
  Ansatz & Orbitals & Optimization level & VMC energy  & DMC energy  \\ 
\midrule \midrule
\multirow{3}*{Single det} & HF & Jas & -97.0284(2) & -97.1041(5)  \\ 

& B3LYP & Jas &  -97.0543(2) & -97.1304(5) \\ 
 
 & opt   & Jas+MO & -97.0570(2) & -97.1306(5)  \\  
 \midrule
  \multirow{3}*{CAS} & RAS2 & Jas & -97.0655(2)  & -97.1318(5) \\ 
 & RAS2	& Jas+CI & -97.0778(2) & -97.1406(5) \\
 & opt    & Jas+MO+CI & -97.0822(3) & -97.1426(4) \\ 
\bottomrule
\end{tabular} 
\end{table}

\end{suppinfo}

\bibliography{library}

\end{document}